\def\(#1){(\call{#1})}
\def\call#1{{#1}}


\font\twelverm=cmr10 scaled 1200    \font\twelvei=cmmi10 scaled 1200
\font\twelvesy=cmsy10 scaled 1200   \font\twelveex=cmex10 scaled 1200
\font\twelvebf=cmbx10 scaled 1200   \font\twelvesl=cmsl10 scaled 1200
\font\twelvett=cmtt10 scaled 1200   \font\twelveit=cmti10 scaled 1200

\skewchar\twelvei='177   \skewchar\twelvesy='60


\def\twelvepoint{\normalbaselineskip=12.4pt
  \abovedisplayskip 12.4pt plus 3pt minus 9pt
  \belowdisplayskip 12.4pt plus 3pt minus 9pt
  \abovedisplayshortskip 0pt plus 3pt
  \belowdisplayshortskip 7.2pt plus 3pt minus 4pt
  \smallskipamount=3.6pt plus1.2pt minus1.2pt
  \medskipamount=7.2pt plus2.4pt minus2.4pt
  \bigskipamount=14.4pt plus4.8pt minus4.8pt
  \def\rm{\fam0\twelverm}          \def\it{\fam\itfam\twelveit}%
  \def\sl{\fam\slfam\twelvesl}     \def\bf{\fam\bffam\twelvebf}%
  \def\mit{\fam 1}                 \def\cal{\fam 2}%
  \def\tt{\twelvett}
  \def\nullspace{\nulldelimiterspace=0pt \mathsurround=0pt }
  \def\big##1{{\hbox{$\left##1\vbox to 10.2pt{}\right.\nullspace$}}}
  \def\Big##1{{\hbox{$\left##1\vbox to 13.8pt{}\right.\nullspace$}}}
  \def\bigg##1{{\hbox{$\left##1\vbox to 17.4pt{}\right.\nullspace$}}}
  \def\Bigg##1{{\hbox{$\left##1\vbox to 21.0pt{}\right.\nullspace$}}}
  \textfont0=\twelverm   \scriptfont0=\tenrm   \scriptscriptfont0=\sevenrm
  \textfont1=\twelvei    \scriptfont1=\teni    \scriptscriptfont1=\seveni
  \textfont2=\twelvesy   \scriptfont2=\tensy   \scriptscriptfont2=\sevensy
  \textfont3=\twelveex   \scriptfont3=\twelveex  \scriptscriptfont3=\twelveex
  \textfont\itfam=\twelveit
  \textfont\slfam=\twelvesl
  \textfont\bffam=\twelvebf \scriptfont\bffam=\tenbf
  \scriptscriptfont\bffam=\sevenbf
  \normalbaselines\rm}



\def\beginlinemode{\endmode
  \begingroup\parskip=0pt \obeylines\def\\{\par}\def\endmode{\par\endgroup}}
\def\beginparmode{\endmode
  \begingroup \def\endmode{\par\endgroup}}
\let\endmode=\par
{\obeylines\gdef\
{}}
\def\singlespace{\baselineskip=\normalbaselineskip}

\def\oneandahalfspace{\baselineskip=\normalbaselineskip
  \multiply\baselineskip by 3 \divide\baselineskip by 2}
\def\doublespace{\baselineskip=\normalbaselineskip \multiply\baselineskip by 2}

\newcount\firstpageno
\firstpageno=2
\footline={\ifnum\pageno<\firstpageno{\hfil}\else{\hfil\twelverm\folio\hfil}\fi}
\let\rawfootnote=\footnote		
\def\footnote#1#2{{\rm\singlespace\parindent=0pt\rawfootnote{#1}{#2}}}
\def\raggedcenter{\leftskip=4em plus 12em \rightskip=\leftskip
  \parindent=0pt \parfillskip=0pt \spaceskip=.3333em \xspaceskip=.5em
  \pretolerance=9999 \tolerance=9999
  \hyphenpenalty=9999 \exhyphenpenalty=9999 }
\def\dateline{\rightline{\ifcase\month\or
  January\or February\or March\or April\or May\or June\or
  July\or August\or September\or October\or November\or December\fi
  \space\number\year}}
\def\received{\vskip 3pt plus 0.2fill
 \centerline{\sl (Received\space\ifcase\month\or
  January\or February\or March\or April\or May\or June\or
  July\or August\or September\or October\or November\or December\fi
  \qquad, \number\year)}}


\hsize=6.5truein
\vsize=8.9truein
\parskip=\medskipamount
\twelvepoint		
\doublespace		
\overfullrule=0pt	



\def\title			
  {\null\vskip 3pt plus 0.2fill
   \beginlinemode \doublespace \raggedcenter \bf}

\def\author			
  {\vskip 3pt plus 0.2fill \beginlinemode
   \singlespace \raggedcenter}

\def\affil			
  {\vskip 3pt plus 0.1fill \beginlinemode
   \oneandahalfspace \raggedcenter \sl}

\def\abstract			
  {\vskip 3pt plus 0.3fill \beginparmode
   \doublespace \narrower ABSTRACT: }

\def\endtitlepage		
  {\endpage			
   \body}

\def\body			
  {\beginparmode}		

\def\head#1{			
  \filbreak\vskip 0.5truein	
  {\immediate\write16{#1}
   \raggedcenter \uppercase{#1}\par}
   \nobreak\vskip 0.25truein\nobreak}

\def\refto#1{$^{#1}$}		

\def\references			
  {\head{References}		
   \beginparmode
   \frenchspacing \parindent=0pt \leftskip=1truecm
   \parskip=8pt plus 3pt \everypar{\hangindent=\parindent}}

\gdef\refis#1{\indent\hbox to 0pt{\hss#1.~}}	

\gdef\journal#1, #2, #3, 1#4#5#6{		
    {\sl #1~}{\bf #2}, #3, (1#4#5#6)}		

\gdef\journ2 #1, #2, #3, 1#4#5#6{		
    {\sl #1~}{\bf #2}: #3, (1#4#5#6)}		

\def\refstylenp{		
  \gdef\refto##1{ [##1]}				
  \gdef\refis##1{\indent\hbox to 0pt{\hss##1)~}}	
  \gdef\journal##1, ##2, ##3, ##4 {			
     {\sl ##1~}{\bf ##2~}(##3) ##4 }}

\def\refstyleprnp{		
  \gdef\refto##1{ [##1]}				
  \gdef\refis##1{\indent\hbox to 0pt{\hss##1)~}}	
  \gdef\journal##1, ##2, ##3, 1##4##5##6{		
    {\sl ##1~}{\bf ##2~}(1##4##5##6) ##3}}

\def\endreferences{\body}

\def\figurecaptions		
  {\endpage
   \beginparmode
   \head{Figure Captions}
}

\def\endpage			
  {\vfill\eject}

\def\endpaper			
  {\endmode\vfill\supereject}

\def\endit
  {\endpaper\end}


\def\ref#1{Ref. #1}			
\def\Ref#1{Ref. #1}			

\def\frac#1#2{{\textstyle #1 \over \textstyle #2}}

\def\half{{\textstyle {1 \over 2}}}
\def\eg{{\it e.g.,\ }}

\def\sla{\raise.15ex\hbox{$/$}\kern-.57em}
\def\leaderfill{\leaders\hbox to 1em{\hss.\hss}\hfill}
\def\twiddle{\lower.9ex\rlap{$\kern-.1em\scriptstyle\sim$}}
\def\bigtwiddle{\lower1.ex\rlap{$\sim$}}
\def\gtwid{\mathrel{\raise.3ex\hbox{$>$\kern-.75em\lower1ex\hbox{$\sim$}}}}
\def\ltwid{\mathrel{\raise.3ex\hbox{$<$\kern-.75em\lower1ex\hbox{$\sim$}}}}
\def\square{\kern1pt\vbox{\hrule height 1.2pt\hbox{\vrule width 1.2pt\hskip 3pt
   \vbox{\vskip 6pt}\hskip 3pt\vrule width 0.6pt}\hrule height 0.6pt}\kern1pt}

\catcode`@=11
\newcount\r@fcount \r@fcount=0
\newcount\r@fcurr
\immediate\newwrite\reffile
\newif\ifr@ffile\r@ffilefalse
\def\w@rnwrite#1{\ifr@ffile\immediate\write\reffile{#1}\fi\message{#1}}

\def\writer@f#1>>{}
\def\referencefile{
  \r@ffiletrue\immediate\openout\reffile=\jobname.ref%
  \def\writer@f##1>>{\ifr@ffile\immediate\write\reffile%
    {\noexpand\refis{##1} = \csname r@fnum##1\endcsname = %
     \expandafter\expandafter\expandafter\strip@t\expandafter%
     \meaning\csname r@ftext\csname r@fnum##1\endcsname\endcsname}\fi}%
  \def\strip@t##1>>{}}

\def\citeall#1{\xdef#1##1{#1{\noexpand\cite{##1}}}}
\def\cite#1{\each@rg\citer@nge{#1}}	

\def\each@rg#1#2{{\let\thecsname=#1\expandafter\first@rg#2,\end,}}
\def\first@rg#1,{\thecsname{#1}\apply@rg}	
\def\apply@rg#1,{\ifx\end#1\let\next=\relax
\else,\thecsname{#1}\let\next=\apply@rg\fi\next}

\def\citer@nge#1{\citedor@nge#1-\end-}	
\def\citer@ngeat#1\end-{#1}
\def\citedor@nge#1-#2-{\ifx\end#2\r@featspace#1 
  \else\citel@@p{#1}{#2}\citer@ngeat\fi}	
\def\citel@@p#1#2{\ifnum#1>#2{\errmessage{Reference range #1-#2\space is bad.}%
    \errhelp{If you cite a series of references by the notation M-N, then M and
    N must be integers, and N must be greater than or equal to M.}}\else%
 {\count0=#1\count1=#2\advance\count1 by1\relax\expandafter\r@fcite\the\count0,%
  \loop\advance\count0 by1\relax
    \ifnum\count0<\count1,\expandafter\r@fcite\the\count0,%
  \repeat}\fi}

\def\r@featspace#1#2 {\r@fcite#1#2,}	
\def\r@fcite#1,{\ifuncit@d{#1}
    \newr@f{#1}%
    \expandafter\gdef\csname r@ftext\number\r@fcount\endcsname%
                     {\message{Reference #1 to be supplied.}%
                      \writer@f#1>>#1 to be supplied.\par}%
 \fi%
 \csname r@fnum#1\endcsname}
\def\ifuncit@d#1{\expandafter\ifx\csname r@fnum#1\endcsname\relax}%
\def\newr@f#1{\global\advance\r@fcount by1%
    \expandafter\xdef\csname r@fnum#1\endcsname{\number\r@fcount}}

\let\r@fis=\refis			
\def\refis#1#2#3\par{\ifuncit@d{#1}
   \newr@f{#1}%
   \w@rnwrite{Reference #1=\number\r@fcount\space is not cited up to now.}\fi%
  \expandafter\gdef\csname r@ftext\csname r@fnum#1\endcsname\endcsname%
  {\writer@f#1>>#2#3\par}}

\def\ignoreuncited{
   \def\refis##1##2##3\par{\ifuncit@d{##1}%
     \else\expandafter\gdef\csname r@ftext\csname r@fnum##1\endcsname\endcsname%
     {\writer@f##1>>##2##3\par}\fi}}

\def\r@ferr{\endreferences\errmessage{I was expecting to see
\noexpand\endreferences before now;  I have inserted it here.}}
\let\r@ferences=\references
\def\references{\r@ferences\def\endmode{\r@ferr\par\endgroup}}

\let\endr@ferences=\endreferences
\def\endreferences{\r@fcurr=0
  {\loop\ifnum\r@fcurr<\r@fcount
    \advance\r@fcurr by 1\relax\expandafter\r@fis\expandafter{\number\r@fcurr}%
    \csname r@ftext\number\r@fcurr\endcsname%
  \repeat}\gdef\r@ferr{}\endr@ferences}


\let\r@fend=\endpaper\gdef\endpaper{\ifr@ffile
\immediate\write16{Cross References written on []\jobname.REF.}\fi\r@fend}

\catcode`@=12

\citeall\refto		
\citeall\ref		%
\citeall\Ref		%
\catcode`@=11
\newcount\tagnumber\tagnumber=0

\immediate\newwrite\eqnfile
\newif\if@qnfile\@qnfilefalse
\def\write@qn#1{}
\def\writenew@qn#1{}
\def\w@rnwrite#1{\write@qn{#1}\message{#1}}
\def\@rrwrite#1{\write@qn{#1}\errmessage{#1}}

\def\taghead#1{\gdef\t@ghead{#1}\global\tagnumber=0}
\def\t@ghead{}

\expandafter\def\csname @qnnum-3\endcsname
  {{\t@ghead\advance\tagnumber by -3\relax\number\tagnumber}}
\expandafter\def\csname @qnnum-2\endcsname
  {{\t@ghead\advance\tagnumber by -2\relax\number\tagnumber}}
\expandafter\def\csname @qnnum-1\endcsname
  {{\t@ghead\advance\tagnumber by -1\relax\number\tagnumber}}
\expandafter\def\csname @qnnum0\endcsname
  {\t@ghead\number\tagnumber}
\expandafter\def\csname @qnnum+1\endcsname
  {{\t@ghead\advance\tagnumber by 1\relax\number\tagnumber}}
\expandafter\def\csname @qnnum+2\endcsname
  {{\t@ghead\advance\tagnumber by 2\relax\number\tagnumber}}
\expandafter\def\csname @qnnum+3\endcsname
  {{\t@ghead\advance\tagnumber by 3\relax\number\tagnumber}}

\def\equationfile{%
  \@qnfiletrue\immediate\openout\eqnfile=\jobname.eqn%
  \def\write@qn##1{\if@qnfile\immediate\write\eqnfile{##1}\fi}
  \def\writenew@qn##1{\if@qnfile\immediate\write\eqnfile
    {\noexpand\tag{##1} = (\t@ghead\number\tagnumber)}\fi}
}

\def\callall#1{\xdef#1##1{#1{\noexpand\call{##1}}}}
\def\call#1{\each@rg\callr@nge{#1}}

\def\each@rg#1#2{{\let\thecsname=#1\expandafter\first@rg#2,\end,}}
\def\first@rg#1,{\thecsname{#1}\apply@rg}
\def\apply@rg#1,{\ifx\end#1\let\next=\relax%
\else,\thecsname{#1}\let\next=\apply@rg\fi\next}

\def\callr@nge#1{\calldor@nge#1-\end-}
\def\callr@ngeat#1\end-{#1}
\def\calldor@nge#1-#2-{\ifx\end#2\@qneatspace#1 %
  \else\calll@@p{#1}{#2}\callr@ngeat\fi}
\def\calll@@p#1#2{\ifnum#1>#2{\@rrwrite{Equation range #1-#2\space is bad.}
\errhelp{If you call a series of equations by the notation M-N, then M and
N must be integers, and N must be greater than or equal to M.}}\else%
 {\count0=#1\count1=#2\advance\count1 by1\relax\expandafter\@qncall\the\count0,%
  \loop\advance\count0 by1\relax%
    \ifnum\count0<\count1,\expandafter\@qncall\the\count0,%
  \repeat}\fi}

\def\@qneatspace#1#2 {\@qncall#1#2,}
\def\@qncall#1,{\ifunc@lled{#1}{\def\next{#1}\ifx\next\empty\else
  \w@rnwrite{Equation number \noexpand\(>>#1<<) has not been defined yet.}
  >>#1<<\fi}\else\csname @qnnum#1\endcsname\fi}

\let\eqnono=\eqno
\def\eqno(#1){\tag#1}
\def\tag#1$${\eqnono(\displayt@g#1 )$$}

\def\aligntag#1$${\gdef\tag##1\\{&(\displayt@g##1 )\cr}\eqalignno{#1\\}$$
  \gdef\tag##1$${\eqnono(\displayt@g##1 )$$}}

\def\eqalignno#1{\displ@y \tabskip\centering
  \halign to\displaywidth{\hfil$\displaystyle{##}$\tabskip\z@skip
    &$\displaystyle{{}##}$\hfil\tabskip\centering
    &\llap{$\displayt@gpar##$}\tabskip\z@skip\crcr
    #1\crcr}}

\def\displayt@gpar(#1){(\displayt@g#1 )}

\def\displayt@g#1 {\rm\ifunc@lled{#1}\global\advance\tagnumber by1
        {\def\next{#1}\ifx\next\empty\else\expandafter
        \xdef\csname @qnnum#1\endcsname{\t@ghead\number\tagnumber}\fi}%
  \writenew@qn{#1}\t@ghead\number\tagnumber\else
        {\edef\next{\t@ghead\number\tagnumber}%
        \expandafter\ifx\csname @qnnum#1\endcsname\next\else
        \w@rnwrite{Equation \noexpand\tag{#1} is a duplicate number.}\fi}%
  \csname @qnnum#1\endcsname\fi}

\def\ifunc@lled#1{\expandafter\ifx\csname @qnnum#1\endcsname\relax}

\let\@qnend=\end\gdef\end{\if@qnfile
\immediate\write16{Equation numbers written on []\jobname.EQN.}\fi\@qnend}

\catcode`@=12


\def\viz{{\it viz.~}}

\def\se{\raise.15ex\hbox{$/$}\kern-.56em\hbox{$\epsilon$}}

\def\figurecaptions{\head{\noindent\bf Figure Captions}
                \beginparmode\frenchspacing
               \leftskip=-.60truecm\parskip=8pt plus 3pt
                \everypar{\hangindent=4em}}

\centerline{\bf The Reduction of the State Vector and Limitations on
Measurement}
\centerline{\bf in the Quantum Mechanics of Closed Systems\footnote{*}
{\sl To appear in the festschrift for Dieter
Brill, edited by B.-L. Hu and T. Jacobson, Cambridge University Press,
Cambridge, 1993}}
\vskip .50 in
\singlespace
\centerline{\sl James B. Hartle\footnote{\S}{e-mail:
hartle@cosmic.physics.ucsb.edu}}
\centerline{\sl Department of Physics}
\centerline{\sl University of California}
\centerline{\sl Santa Barbara, CA 93106}
\vskip .5in
{\hfil \it ``... persuaded of these principles, what havoc must we
make?''} --
Hume
\vskip .5in
\centerline{\it ABSTRACT}
\vskip .26 in
Measurement is a fundamental notion in the usual
approximate quantum mechanics of measured subsystems.
Probabilities are predicted for the outcomes of measurements.
State vectors evolve unitarily in between measurements and by reduction
of the state vector at measurements. Probabilities are computed by
summing the squares of amplitudes over alternatives which could have
been measured but weren't. Measurements are limited by
uncertainty principles and by other restrictions arising from the
principles of quantum mechanics. 
This essay examines
the extent to which those features of the quantum mechanics of measured
subsystems that are explicitly 
tied to measurement situations are incorporated
or modified in the more general quantum mechanics of closed systems
in which measurement is not a fundamental notion. There, probabilities
are predicted for decohering sets of alternative time
histories
of the closed system, whether or not they represent a measurement
situation. Reduction of the state vector is a necessary part of the
description of such histories.
Uncertainty principles limit the possible alternatives at one
time from which histories may be constructed.
Models of
measurement situations are exhibited within the quantum mechanics of the
closed system containing both measured subsystem and measuring
apparatus.  Limitations are derived
 on the existence of records for the outcomes
of measurements when the initial density matrix of the closed
system is highly impure.
\vfill
\eject
\doublespace
\centerline{\bf 0. Preface}
\taghead{0.}

In 1959, then an undergraduate at Princeton in search of a senior thesis
topic, I was introduced by John Wheeler to his young colleague, Dieter
Brill. This was fortunate from my point of view, for Dieter
proved to have the patience, time and talent not only to introduce me to the 
beauties of Einstein's general relativity but also give me instruction and 
guidance in the
practice of research. Our subject -- the method of the
self-consistent field in general relativity and its application to the
gravitational geon
-- was also fortunate. Through it we helped lay the foundations for 
the short wavelength approximation for gravitational radiation (Brill and
Hartle, 1964). In
particular, building on ideas of Wheeler (1964), we introduced what Richard Isaacson (Isaacson 1968ab) was
later kind enough to call the ``Brill-Hartle'' average for the effective
stress-energy tensor of short wavelength radiation, and which was to to
prove such a powerful tool when made precise in his hands in his general
theory of this approximation.
It would be difficult to imagine a more marvelous introduction to
research. 
I have not written a paper with Dieter Brill since, but each day I use 
the lessons learned from him so long ago. 
It is a pleasure to thank him with this small essay on the occasion of
his 60th birthday.

\vskip .26in
\taghead{I.}
\centerline{\bf I. Introduction}

``Measurement'' is central to the usual formulations of quantum
mechanics. Probabilities are predicted for the outcomes of measurements
carried out on some subsystems of the universe by others. In a
Hamiltonian formulation of quantum mechanics, states of a subsystem
evolve unitarily in between measurements and by reduction of the state
vector at them. In a sum-over-histories formulation, amplitudes are
squared and summed over alternatives ``which could have been measured
but weren't'' to calculate the probabilities of incomplete measurements.
In these and other ways the notion of measurement plays a fundamental
role in the usual formulations of quantum theory.

The quantum mechanics of a subsystem alone, of course, does not offer
a quantum mechanical description of the workings of the measuring
apparatus which acts upon it, but it does limit what can be measured. 
We cannot, for instance,  carry out simultaneous ideal measurements of 
the position and momentum of a particle to arbitrary accuracies.  Ideal
measurements are defined to
 leave the subsystem in eigenstates of the measured
quantities and there are no states of the subsystem for which position
and momentum are specified to accuracies better than those allowed by
the Heisenberg uncertainty principle.  Analyses of the workings of
measuring apparatus  and subsystem as part of a single quantum system
reveal further quantum mechanical limitations on ideal measurements, as
in the work of Wigner (1952) and Araki and Yanase (1960).

Cosmology is one motivation for generalizing the quantum
mechanics of measured subsystems to a quantum mechanics of closed
systems in which measurement plays no fundamental role. Simply providing
a more coherent and precise formulation of quantum mechanics, free from
many of the usual interpretive difficulties, is motivation enough for
many. Today, because of the efforts of many over the last thirty-five
years, we have a quantum mechanics of closed 
systems.\footnote{$^*$}{A pedagogical introduction to the 
quantum mechanics of closed
systems can be found in the author's other contribution to these
volumes, (Hartle, 1993a), where references to some of the literature may
be found.}
In this formulation, it is the internal consistency of probability sum
rules that determines the sets of alternatives of the closed system for
which probabilities are predicted rather than any external notion of 
``measurement'' (Griffiths, 1984; Omn\`es, 1988abc; Gell-Mann and
Hartle, 1990). It is the absence of quantum mechanical interference
between the individual members of a set of alternatives, or decoherence,
that is a sufficient condition for the consistency of probability sum
rules. It is the initial condition of the closed system that, together
with its Hamiltonian, determines
which sets of alternatives decohere and which do not. Alternatives
describing a measurement situation decohere, but an alternative does not
have to be part of a measurement situation in order to decohere. Thus,
for example, with an initial condition and Hamiltonian are such that they
decohere, probabilities are predicted for alternative sizes of density
fluctuations in the early universe or alternative positions of the moon
whether or not they are ever measured.

The familiar quantum mechanics of measured subsystems is an
approximation to this more general quantum mechanics of closed systems.
It is an approximation that is appropriate when certain approximate
features of measurement situations can be idealized as exact. These
include the decoherence of alternative configurations of the apparatus
in which the result of the measurement is registered, the correlation of
these with the measured alternatives,  the short 
duration of certain measurement interactions compared to characteristic
dynamical time scales of the measured subsystems, the persistence of the
records of measurements, 
etc.~etc.\footnote{$^*$}{For more discussion of ideal measurement 
models in the
context of the quantum mechanics of closed systems see Section IV and
Hartle (1991a)}
The question naturally arises as to the extent to
which those features of the quantum mechanics of measured subsystems
that were tied to measurement situations are incorporated, modified, or
dispensed with in the more general quantum mechanics of closed systems.
Are two laws of evolution still needed? Is there reduction of the state
vector, and if so, when? What becomes of a rule like ``square amplitudes
and sum over probabilities that one could have measured but didn't''?
What becomes of the limitations on measurements in a more general theory
where measurement can be described but does not play a fundamental role.
This essay is devoted to some thoughts on these questions.

\vskip .26in
\taghead{2.}
\centerline{\bf II. The Reduction of the State Vector}
\vskip .13in 
In the approximate quantum mechanics of measured subsystems the
Schr\"odinger picture state of the subsystem is
described by a time-dependent
 vector, $|\psi(t)\rangle$, in the subsystem's Hilbert space.
In between measurements the state vector evolves unitarily:
$$ 
  i\hbar \frac{\partial\vert\psi(t)\rangle}{\partial t} = 
h \vert\psi(t)\rangle\ . \tag twoone
$$
If a measurement is carried out at time $t_k$, the probabilities for its
outcomes are
$$
p(\alpha_k) = \bigl\Vert s^k_{\alpha_k} | \psi (t_k) \rangle
\bigr\Vert^2\ . \tag twotwo
$$
Here, the $\{s^k_{\alpha_k}\}$ are 
an exhaustive set of orthogonal, Schr\"odinger picture,
projection operators describing the
possible outcomes. The index $k$ denotes the {\it set} of outcomes at
time $t_k$, for example, a set of ranges of momentum, or a set of ranges
of position, etc. The index $\alpha_k$ denotes the particular
alternative within the set -- a {\it particular} range of momentum, a
{\it particular} range of position, etc. If the measurement was an
``ideal'' one, that ``disturbed the system as little as possible'', the
state vector is reduced at $t_k$ by the projection that describes the
outcome of the measurement:
$$
\vert\psi(t_k)\rangle\to\frac{s^k_{\alpha_k}\vert\psi(t_k)\rangle}{\Vert
s^k_{\alpha_k}\vert\psi(t_k)\rangle\Vert}\ . \tag twothree
$$
This is the ``second law of evolution'',  which together with the first
\(twoone), can be used to calculate the probabilities of sequences of ideal
measurements.

The two laws of evolution can be given a more unified expression. For
example, in the Heisenberg picture, the joint probability of a sequence
of measured outcomes is given by the single expression\footnote{$^*$}{The
utility of the Heisenberg picture in giving a compact expression for the
two laws of evolution has been noted by many authors, 
Groenewold (1952) and Wigner (1963), among the earliest.
Similar unified expressions can be
given in the sum-over-histories formulation of quantum mechanics (Caves
1986, 1987 and Stachel 1986)}:
$$
p(\alpha_n,\dots,\alpha_1)=\Vert s^n_{\alpha_n}(t_n)\cdots s^1_{\alpha_1}(t_1)
\vert\psi\rangle\Vert^2 \tag twofour
$$
where $|\psi\rangle$ is the Heisenberg state vector and
$$
s^k_{\alpha_k}(t_k) = e^{iht_k/\hbar}s^k_{\alpha_k}e^{-iht_k/\hbar}
\tag twofive
$$
are the Heisenberg picture projection operators with $h$ the Hamiltonian
of the subsystem.
Nevertheless, even in such compact expressions one can distinguish
unitary evolution from the action of projections at an ``ideal"
measurement.

One gains the impression from parts of the literature that some think
the law of state vector reduction to be secondary in importance
to the law of unitary evolution. Perhaps by understanding the
quantum mechanics of large, ``macroscopic'' systems that include the
measuring apparatus the second law of
evolution can be derived from the first.  Perhaps the law of the
reduction of the state vector is unimportant for the calculation of
realistic probabilities of physical interest. No ideas could be further
from the truth in this author's opinion. Certainly the second law of 
evolution is less precisely formulated that the law of unitary evolution
because the notion of an ``ideal'' measurement is vague and many
realistic measurements are not very ideal. However, as shown
conclusively by Wigner (1963), the second law of
evolution is not reducible to the first and it is essential for the
calculation of probabilities of realistic, everyday interest as we shall
now describe.

Scattering experiments can perhaps be said to involve but a single
measurement of the final state once the system has been prepared in an
initial state. Many everyday probabilities, however are for {\it time
sequences} of measurements. For instance, in asserting that the moon
moves on a certain classical orbit one is asserting that successions of
suitably crude measurements of the moon's position and momentum will be
correlated in time by Newton's deterministic law. Thus, measured classical
behavior involves probabilities for
 time sequences like \(twofour). Successive
state vector reductions are essential for their prediction as well as
many other questions of interest in quantum
mechanics.

Since the state vector of a subsystem evolves unitarily except when that
subsystem is measured by an external device, some have argued that 
one could dispense with the
``second law of evolution''  in
the quantum mechanics of a closed system. All predictions would be
derived from a state
vector, $|\Psi(t)\rangle$, of the closed system that evolves in time
only
according to the Schr\"odinger equation  (Everett, 1957; DeWitt, 1970).
However, a state vector is a function of one time and can, therefore, be
used to predict only the probabilities of alternatives that are at one time
according to the generalization of \(twotwo)
$$
p(\alpha_k) = \bigl\Vert P^k_{\alpha_k} | \Psi (t_k) \rangle
\bigr\Vert^2\ . \tag{twosix}
$$
Here, the $\{P^k_{\alpha_k}\}$ are an exhaustive set of orthogonal,
Schr\"odinger picture, projection operators representing alternatives of
the closed system at a moment of time. For instance, in a description of
the system in terms of hydrodynamic variables they might represent
alternative ranges of the energy density averaged over suitable volumes.
In a description of a measurement situation, the $P's$ might represent
alternative registrations of that variable by an apparatus. 

The restriction to a unitary law of evolution and the action of
projections at a single time as in \(twosix) would rule out the
calculation of probabilities for time histories of the closed system.
Some have suggested that probabilities at the single marvelous moment of
time ``now'' are enough for all realistic physical prediction and
retrodiction.\footnote{$^*$}{For a recent expression of this point of
view, see Page and Wootters (1983).} In this view, for example, 
probabilities referring to past
history  are more realistically understood as
the probabilities for correlations among present records.  However, just
to establish whether a physical system is a good record, one needs to
examine the probability for the correlations between the present value
of that record and the past event it has supposed to have recorded.  That
is a probability for correlation between alternatives at two times ---
the probability of a history.  For this and other reasons, probabilities
of histories are just as essential in the quantum mechanics of closed systems
as they were in the quantum mechanics of measured subsystems.

There is a natural generalization of expressions like \(twofour) to
give a framework for predicting the joint probabilities of time
sequences of alternatives in the quantum mechanics of closed systems
(Griffiths, 1984; Omn\`es, 1988abc; Gell-Mann and Hartle, 1990).
The joint probability of a history of alternatives is
$$           
p(\alpha_n,\dots,\alpha_1)=\Vert P^n_{\alpha_n}(t_n)
\cdots P^1_{\alpha_1}(t_1)\vert
\Psi\rangle\Vert^2 \tag twoseven
$$
where the Heisenberg $P$'s evolve according to 
$$
P^k_{\alpha_k}(t_k) = e^{iHt_k/\hbar}P^k_{\alpha_k}(0)\ e^{-iHt_k/\hbar}
\ .\tag twoeight
$$
and the times in \(twoseven) are ordered with the earliest closed to
$|\Psi\rangle$.
Here, projection operators, state vectors, the Hamiltonian $H$, etc
all refer to the Hilbert space of a closed system, containing both
apparatus and measured subsystem if any.  This is most generally the
universe, in an approximation in which gross quantum
fluctuations in the geometry of spacetime can be 
neglected.\footnote{$^*$}{For
a generalized quantum mechanics of closed systems that includes quantum
spacetime see Hartle (1993b) and the references therein.} The Heisenberg
state vector
$|\Psi \rangle$ represents the initial condition 
of the closed system, assumed here to be a pure state
for simplicity.  

The occurrence of the projections in \(twoseven) can be described by
saying that the state vector is ``reduced'' at each instant of time
where an alternative is considered.  However,
the important point for the present discussion of state vector reduction
is that the projections in \(twoseven) are not, perforce, associated with a 
measurement by some external system. This is a quantum mechanics of a
closed system! The $P's$ can represent any alternative at a moment of
time. Measurement situations within a closed system of apparatus and
measured subsystem can be described by appropriate $P's$ (see Section
IV) but the $P's$ do not necessarily have to describe measurement
situations. They might describe alternative positions of the moon
whether or not it is being observed or alternative values of density
fluctuations in the early universe where ordinary measurement situations
of any kind are unlikely to have existed.  Thus, the state vector can be said
to be reduced in \(twoseven) by the action of the projections and one
might even say that there are ``two laws of evolution'' present, but
those reductions and evolutions have nothing to do, in general, with
measurement situations.  In the author's view, it is clearer not to use
the language of ``reduction'' and ``two laws of evolution'',
 but simply to regard \(twoseven) as the law for the joint
probability of a sequence of alternatives of a closed system.
 Projections
occur therein because they are the way alternatives are represented in
the quantum mechanics of closed systems.

There is a good reason why the probabilities \(twofour) of a sequence of
alternatives of a subsystem refer only to
the results of {\it measured} alternatives. 
It would be inconsistent generally to calculate probabilities
of histories that have not been measured because the sum rules of
probability theory would not be satisfied as a consequence of quantum
mechanical interference. In the two-slit experiment,
for instance, the probability to arrive at a point on the screen is not
the sum of the probabilities to go through the alternative slits and 
arrive at that point unless the alternative passages have been measured
and the interference between them destroyed.
Thus, probabilities are not predicted for all possible sets of histories
of a subsystem but only those which have been ``measured''.

Probabilities are not predicted for every set of alternative histories
of a closed system either. But it is not an external notion of
``measured'' that discriminates those sets for which probabilities are 
predicted from those which are not. Rather, it is the internal
consistency of the probability sum rules that distinguishes them
(Griffiths, 1984). Probabilities are consistent for a set of histories,
when, in a partition of the set of histories into an exhaustive set of 
exclusive classes, the probabilities of the individual classes are the
sums of the probabilities of the histories they contain for all allowed
partitions. A sufficient condition for the consistency of probabilities
is
the absence of interference between the individual histories in the set
as measured by the overlap
$$
\langle\Psi\vert P^n_{\alpha'_n}(t_n)\cdots P^1_{\alpha'_1}(t_1) \cdot
P^1_{\alpha_1}(t_1)\cdots P^n_{\alpha_n}(t_n)\vert\Psi\rangle \propto
\delta_{\alpha'_1 \alpha_1}\cdots\delta_{\alpha'_n\alpha_n}.
\tag twonine
$$
Sets of histories that satisfy \(twonine) are said to {\it
decohere}.\footnote{$^*$}{There are several possible decoherence conditions.
This is {\it medium decoherence} in the terminology of Gell-Mann and
Hartle (1990b).} Decoherence implies the consistency of the probability
sum rules. In the quantum mechanics of closed systems, probabilities are
predicted for just those sets of alternative histories that decohere
according to \(twonine) as a consequence of the system's Hamiltonian
and initial quantum state $\vert\Psi\rangle$.

\vskip .26in
\taghead{3.}
\centerline{\bf III.  Uncertainty Principles}
\vskip .13in
The state of a single particle cannot be simultaneously an eigenstate of
position and momentum. It follows from their commutation relations
that position and momentum cannot be specified to accuracies greater
than those allowed by the Heisenberg uncertainty principle
$$
\Delta x \Delta p \ge \half \hbar \ .   \tag threetwo
$$ 
Following the standard discussion, we infer from the mathematical
inequality \(threetwo) that it is not possible to simultaneously perform
ideal measurements of position and momentum to accuracies better than
that allowed by the uncertainty principle \(threetwo). There can be no
such ideal measurement because there is no projection operator, $s$, that
could represent its outcome in \(twothree). 

The limitations on ideal measurements implied by the uncertainty
principle \(twothree) are usually argued to extend to non-ideal
measurements as well.  Examination of quantum mechanical models of
specific measurement situations have for the most part verified the
consistency of this extension although some have maintained
otherwise.\footnote{$^*$}{For example, Margenau (1958) and Prugove\v cki
(1967).  We are not discussing here, nor do we discuss later,
``unsharp'' observables or ``effects''.  For those see \eg Busch
(1987).} No such elaborate analysis is needed to demonstrate the
impossibility of ideal measurements of position and momentum to
accuracies better than those allowed by the uncertainty principle.  That
limitation follows from the quantum mechanics of the subsystem alone.

The mathematical derivation of the uncertainty relation \(threetwo) is,
of course, no less valid in the Hilbert space of a closed system than it
is for that of a subsystem. In the quantum mechanics of a closed system,
however, the absence of projection operators that specify $x$ and $p$
to accuracies better than \(threetwo) is not to be interpreted as a 
limitation on external measurements of this closed system. By hypothesis
there are none! Rather, in the quantum mechanics of  closed systems,
uncertainty relations like \(threetwo) are limitations on how a closed
system can be {\it described}. There are no histories in which position
and momentum can be simultaneously
specified to accuracies better than allowed by
Heisenberg's principle.

 Although there are no projection operators that specify position and
momentum {\it simultaneously} to accuracies better than the limitations
of the uncertainty principle, we can consider histories in which
position is specified sharply at one time and momentum at another time.
Let $\{\Delta_\alpha\}$ be an exhaustive set of exclusive position
intervals, $\{\tilde\Delta_\beta\}$ be an exhaustive set of exclusive
momentum intervals, and $\{P_\alpha(t')\}$ and $ \{\tilde P_\beta (t)\}$
be the corresponding Heisenberg picture projection operators at times
$t'$ and $t$ respectively. 
An individual history in which the momentum lies in the interval $\tilde
\Delta_\beta$ at time $t$ and the position in interval $\Delta_\alpha$
at a {\it later} time $t'$ would correspond to a branch of the initial 
state vector of the form:
$$
P_\alpha(t')\tilde P_\beta(t)\vert \Psi \rangle\ . \tag threethree
$$
As $\alpha$ and $\beta$ range over all values, an exhaustive set of 
alternative histories of the closed system is generated. Probabilities
are assigned to these histories when the set decoheres, that is, when
the branches \(threethree) are sufficiently orthogonal according to 
\(twonine). 

Nothing prevents us from considering the case when $t'$ coincides with
$t$. If the alternative histories decohere, one would predict the joint
probability $p(\alpha, \beta)$ that the momentum is in the interval
$\tilde\Delta_\beta$ at one time and {\it immediately} afterwards
 the position
is
in the interval $\Delta_\alpha$. That would give a different meaning to
the probability of a simultaneous specification of position and
momentum. 

Even if the intervals $\{\Delta_\alpha\}$ and $\{\tilde\Delta_\beta\}$
are infinitesimal, corresponding to a sharp specification of position
and momentum there are some states $\vert\Psi\rangle$ for which these
alternatives decohere. Eigenstates of momentum provide one example. 
However, for no state $\vert\Psi\rangle$  will the marginal
probability distributions of position and momentum have variances that
violate the uncertainty principle. That is because decoherence implies
the probability sum rules so that
$$
p(\alpha)\equiv\sum_\beta p(\alpha,\beta) = \Vert P_\alpha (t)\vert\Psi
\rangle\Vert^2  \tag threefour a
$$
and
$$
p(\beta)\equiv\sum_\alpha p(\alpha,\beta) = \Vert \tilde P_\beta (t)\vert\Psi
\rangle\Vert^2  \tag threefour b
$$
where, in each case, the last equality follows from decoherence.
However, the left-hand sides of \(threefour)
are
just the usual  probabilities for position and momentum computed from a
single state. Their variances must satisfy the uncertainty principle. 

One would come closer to the classical meaning of simultaneously
specifying the position and momentum if histories of coincident position
and momentum projections decohered independently of their order.  That
is, if the set of histories 
$$
\widetilde P_\beta (t) P_\alpha (t) | \Psi \rangle \tag threefive
$$
were to decohere 
in addition to the set defined by \(threethree) with $t^\prime = t$. 
In that case it is straightforward to show that the
joint probabilities $p(\alpha, \beta)$ are independent of the order
of the projections as a
consequence of decoherence.  Whether states can be exhibited in which
both \(threethree) and \(threefive) decohere is a more difficult question.

\vskip .26in
\taghead{4.}
\centerline{\bf IV. Limitations on Ideal Measurements}
\vskip .13in

While measurement is not fundamental to a formulation of the quantum
mechanics of a closed system, measurement situations can be described
within it. That is because we can always consider a closed system
consisting of measuring apparatus and measured subsystem or most
generally and accurately the entire universe. Roughly speaking, a
{\it measurement situation} is one in which a variable of the measured 
subsystem, perhaps not normally decohering, becomes correlated with high
probability with a variable of the apparatus that decoheres because of
{\it its} interactions with the rest of the universe. The variable of the
apparatus is called a {\it record} of the measurement outcome. The
decoherence of the alternative values of this record leads to the
decoherence of the measured alternatives because of their
correlation. Measurement situations can be described quantitatively
in the quantum mechanics of closed systems by using the overlap
\(twonine) to determine when measured alternatives decohere
and using
the resulting probabilities to assess the degree of correlation between
record and measured variable. By such means, any measurement
situation, ideal or otherwise,
 may be accurately handled in the quantum mechanics of the
closed system containing both measuring apparatus and measured
subsystem.   

Conventional discussions of measurement in quantum mechanics often focus
on {\it ideal measurement models} in which certain approximate features
of realistic measurement situations are idealized as 
exact.\footnote{$^*$}{Some
classic references are von Neumann (1932), London and Bauer (1939),
and Wigner (1963) or see almost any text on quantum mechanics.} In
particular an ideal measurement is one that leaves a subsystem that is
initially in an eigenstate of a measured quantity in that same
eigenstate after the measurement. The subsystem is thus ``disturbed as
little as possible'' by its interaction with the apparatus. Of course,
not very many realistic measurements are ideal in this sense. Typically,
after a measurement, subsystem and apparatus are not even in a product
state for which it makes sense to talk about the ``state of the
subsystem''. Probably the reason for the focus on ideal measurement
models is that they are models of the sorts of measurements for which
the ``reduction of the state vector'' could accurately model the
evolution of the measured subsystem interacting with the measuring
apparatus. In particular, a reduction of the state of the apparatus will
leave the subsystem in the correlated eigenstate of the measured
variable.

Quantum mechanics severely restricts the possible ideal measurement
situations. Wigner (1952) and Araki and Yanase (1960) showed that, even
given arbitrary latitude in the choice of Hamiltonian describing the
combined system of apparatus and measured subsystem, only quantities
that commute with additive, conserved quantities can be ideally
measured. This is a very restrictive conclusion. It rules out, for
example, precise,
ideal measurements of the position and momentum of a particle
with a realistic Hamiltonian. (They do not commute with the additive,
conserved angular momentum.) Araki and Yanase showed that, in a certain
sense, ideal measurements were {\it approximately}
possible for such quantities, but in a strict sense quantum mechanics
prohibits them. 

Impurity of the initial state of a closed system limits ideal
measurements in another way. To derive this limitation it is necessary
to discuss more precisely ideal measurement models in
the quantum mechanics of closed systems.\footnote{$^*$}{For more detail than
can be offered here see Hartle (1991a), Section II.10.}  We will use the
more of the formulation of the quantum mechanics of closed systems than
has been developed here.  The reader can find the necessary background
in the author's other contribution to these volumes, (Hartle 1993a).

We consider a closed system in which we can identify alternatives of a 
subsystem that are to be ``measured''. Let $\{S^k_{\alpha_k}(t_k)\}$,
$\alpha_k=1,2,3,\cdots$ be the Heisenberg projection operators
corresponding to these alternatives at a set of times $\{t_k\}$, $k=1,
2, 3 \cdots$. In a
more detailed ideal measurement model
 we might assume that
the Hilbert space of the closed system is a tensor product of a Hilbert
space ${\cal H}^s$ defining the subsystem and a Hilbert space ${\cal
H}^r$ defining the rest of the universe outside the subsystem. In the
Schr\"odinger picture, projection operators representing the
alternatives of the subsystem would have the form
$S^k_{\alpha_k}=s^k_{\alpha_k} \otimes I^r$  where the $s's$ act on 
${\cal H}^s$ alone. However, such specificity is not needed for the
result that we shall derive. 

Let us consider how a sequence of ideal measurements of alternatives of
the subsystem $S^k_{\alpha_k}(t_k)$ at times $t_1<\cdots <t_n$ is 
described. A history of specific alternatives
$(\alpha_1,\cdots,\alpha_n)\equiv\alpha$ is represented by the
corresponding chain of projections:
$$
C_\alpha=S^n_{\alpha_n}(t_n)\cdots S^1_{\alpha_1}(t_1)\ .  \tag fourone
$$

One defining feature of an ideal measurement situation is that
there should exist at a time $T>t_n$ a {\it record} of the outcomes of
the measurement that is exactly correlated with the measured
alternatives of the subsystem. That is , there should be a set of
orthogonal, 
{\it commuting}, projection operators $\{R_\beta(T)\}$ with
$R_\beta (T)\equiv R_{\beta_n\cdots\beta_1} (T)$ which are always
exactly correlated with the measured alternatives $C_\alpha$ in
histories that contain them both, as a consequence of the system's
initial condition.  The degree of correlation is defined by the
decoherence functional $D(\beta^\prime\alpha^\prime; \beta\alpha)$ which
measures the interference between a history consisting of a sequence of
measured alterhatives $\alpha = (\alpha_1, \cdots, \alpha_n)$ followed
by a record $\beta = (\beta_1, \cdots, \beta_n)$ at time $T$ and a
similar history with alternatives $\alpha^\prime$ and $\beta^\prime$. If
the records are exactly correlated with the measured alternatives.
$$
D\left(\beta^\prime, \alpha^\prime; \beta, \alpha \right) = 
Tr[R_{\beta^\prime}(T)C_{\alpha^\prime}\rho {C^\dagger}_\alpha R_\beta (T)]
\propto \delta_{\beta^\prime \alpha^\prime}\delta_{\beta\alpha} \tag
fourtwo
$$
where $\rho$ is the Heisenberg picture initial density matrix of the
closed system of apparatus and measured subsystem and
$\delta_{\beta\alpha}$ means $\delta_{\beta_1\alpha_1}
\delta_{\beta_2\alpha_2}\cdots \delta_{\beta_n\alpha_n}$, etc. 

The existence of exactly correlated records as described by \(fourtwo)
ensures the decoherence of the histories of the subsystem and permits
the prediction of their probabilities. That is because the records are 
orthogonal and exhaustive:
$$
R_\beta(T) R_{\beta'}(T) = \delta_{\beta\beta'} R_\beta(T), \qquad 
\sum\nolimits_\beta
R_\beta(T)= I\ .  \tag fourthree
$$
These properties together with the cyclic property of the trace are
enough to show that
$$
Tr[C_{\alpha'}\rho {C^\dagger}_\alpha] \propto \delta_{{\alpha'}\alpha}
\tag fourfour
$$ 
follows from \(fourtwo). This is the generalization of the decoherence
condition \(twonine) for an initial density matrix.
 The measurement correlation thus effects the 
decoherence of the measured alternatives. 

Of course, much more is usually demanded of an ideal measurement
situation than just decoherence of the measured alternatives. There is
the idea that ideal measurements ``disturb the measured subsystem as
little as possible'' and, in particular,  that values of measured
quantities are not disturbed. These are described in more detail in the
context of the quantum mechanics of closed systems in Hartle (1991a). 
For our discussion, however, we need only the feature that ideal
measurements assume {\it exactly} correlated records of measurement 
outcomes, for we shall now show that if the density matrix is highly
impure such records cannot exist for non-trivial sets of measured
histories.\footnote{$^*$}{ The argument we shall give is a straightforward
extension of that used by M. Gell-Mann and the author to analyze the 
possibility of ``strong decoherence'' in Gell-Mann and Hartle (1993a). 
Thanks are due to M. Gell-Mann for permission to publish here what is 
essentially a joint result.}

We begin by introducing bases of complete sets of states in which the
density matrix $\rho$ and the commuting set of projection operators
$\{R_\beta(T)\}$ are diagonal, \viz :
$$
\eqalignno{
\rho & = \sum\nolimits_r | r\rangle \pi_r \langle r |\ ,& (fourfive)\cr
R_\beta(T) & = \sum\nolimits_n |\beta, n\rangle\langle \beta, n|\ .
&(foursix)\cr
}
$$
where $\pi_r$ are the diagonal elements of $\rho$.  When ``diagonal''
elements of the condition \(fourtwo) (those with $\alpha =
\alpha^\prime, \beta = \beta^\prime)$ are written out in terms of these
bases they take the form
$$
\sum\nolimits_{r,n} \pi_r | \langle\beta, n| C_\alpha | r \rangle |^2
\propto \delta_{\alpha\beta}\ . \tag fourseven
$$
The left-hand side is a sum of positive numbers so that this implies
$$
\langle r|C_\alpha | \beta, n\rangle = 0\ ,\ {\rm when}\ \alpha\not=
\beta\ , \tag foureight
$$
for all $r$ for which $\pi_r \not= 0$.

If the density matrix $\rho$ is highly impure, so that $\pi_r \not=0$ for
a {\it complete} set of states $\{|r\rangle\}$, the relation
\(foureight) implies the operator condition
$$
C_\alpha | \beta, n\rangle = 0\ ,\quad \alpha \not= \beta\ .
\tag fournine
$$
Therefore, $C_\alpha$ is non-zero only on the subspace defined by $R_\alpha(T)$
where $R_\alpha(T)$ is effectively unity.  Thus we have
$$
R_\beta(T) C_\alpha = \delta_{\beta\alpha} C_\alpha\ .
\tag foureleven
$$
Summing this relation over $C_\alpha$ and utilizing the fact that
$\sum_\alpha C_\alpha = I$, we find
$$
C_\alpha = R_\alpha(T)\ .
\tag fourtwelve
$$
which says that the string of projections is itself a projection.  This
can happen only if the string consists of a single projection or if all
the projections in the string commute
 with each other.  To see the latter fact write
\(fourtwelve) in detail as
$$
S^n_{\alpha_n} (t_n) \cdots S^1_{\alpha_1} (t_1) = R_{\alpha_n \cdots
\alpha_1}(T)\ . \tag fourthirteen
$$
Summation implies
$$
S^k_{\alpha_k} (t_k) = \sum_{\alpha_j\not= \alpha_k} R_{\alpha_n \cdots
\alpha_1}(T)
\tag fourfourteen
$$
but since the $R$'s commute with each other
 the $S$'s must also.  Even in the case
that $C_\alpha$ consists of a single projection, \(fourtwelve) shows that
record and projection are indentical.  If $C_\alpha$ consists of
projections that refer to a subsystem defined by a Hilbert space as
described above then the records cannot be elsewhere in the universe.
{\it Thus, if the initial density matrix 
is highly impure, in the sense that it has non-zero probabilities for a
complete set of states, there cannot be exactly correlated records of
measurement outcomes.  In particular there cannot be ideal measurements.}

Of course, in realistic measurement situations we do not expect to find
records that are {\it exactly} correlated with measured variables of a
subsystem. Neither do we necessarily expect {\it exact} decoherence of
measured alternatives
 or many of the other idealizations of the ideal measurement
situation as very experimentalist knows! It, therefore, becomes an
interesting question to investigate quantitatively the connection
between the $\{\pi_r\}$ of the density matrix and the degree to which
approximate records defined by a relaxed \(fourtwo) exist.

\vskip .26in
\taghead{5.}
\centerline{\bf V. Interfering Alternatives}
\vskip .13in
The starting point for Feynman's sum-over-histories formulation of
quantum mechanics is the prescription of the amplitude for an elementary
(completely fine-grained) history of a measured subsystem as 
$$
\exp[iS({\rm history})/\hbar] \tag fiveone
$$
 where $S$ is the action functional summarizing the subsystem's
dynamics. As an example, we may think of a non-relativistic particle
moving in one dimension. In this case the elementary histories are the
possible paths of the particle, $x(t)$, and the action is the usual
$$
S[x(\tau)]=\int\nolimits dt \left[ \half m \left({dx \over dt}\right)^2 
- V(x) \right]\ .
\tag fivetwo
$$
We will use this example for all illustrative purposes in what follows.

A given experimental situation determines some parts of the
subsystem's path  but leaves undetermined many other parts. For instance, 
consider a measurement that determines whether or not a particle is in
a position interval $\Delta$ at time $t$. In that case the measurement 
leaves undetermined the positions at times other than $t$ and the 
relative position within $\Delta$ at time $t$. Given an initial
state at time $t_0$ represented by a wave function $\psi(x_0)$, we may
compute the probabilities for the outcomes that are
determined by the measurement
as follows: We first divide the {\it undetermined} alternatives into 
``interfering''  and ``non-interfering''
(or ``exclusive'')
alternatives according to the experimental situation. We sum amplitudes
for histories weighted by the initial wave function over the interfering
alternatives, square that, and sum the square over the non-interfering 
alternatives. The result is the probability for the measured
determination.  For example, in the case of the measurement mentioned
above that localized a particle to an interval $\Delta$ at time $t$,
the probability of this outcome is:
$$
p(\Delta)=\int_\Delta d x_f
\left\vert \int_{x_f} \delta x\ e^{iS[x(\tau)]/\hbar}
\psi(x_0) \right\vert^2\ .      \tag fivethree
$$
The path integral is over all paths in the time interval $[t_0,t]$ that
end in $x_f$, and includes an integral over the initial position $x_0$. These
are the ``interfering alternatives''. The square of the amplitude is
summed over the final position within $\Delta$. These positions are the
``non-interfering'' alternatives. 

What determines whether an undetermined alternative is interfering or
not? Certainly it is not whether it is measured in the experimental
situation. In the above example, positions at times other than $t$ were
not measured and they were ``interfering''. But the experiment also did
not measure the relative position within $\Delta$ and this was
``non-interfering''. According to Feynman and Hibbs (1965):
\itemitem{}``It is not hard, with a little experience, to tell what kind
of alternatives is involved. For example, suppose that information
about alternatives is available ({\it or could be made available without
altering the result}) [author's italics], but this information is not 
used. Nevertheless, in this case a sum of probabilities (in the ordinary
sense) must be carried out over {\it exclusive} alternatives. These
exclusive alternatives are those which {\it could} have been separately
identified by the information.''

\noindent Thus, in the above example, the value of $x$ at a 
time other than $t$ is
an interfering alternative because we could not have acquired
information about it without disturbing the later probability
that $x$ is in $\Delta$ at $t$.
By contrast, the precise
 value of $x$ within $\Delta$ is a non-interfering 
or exclusive alternative because we {\it could} have measured it
precisely and left the probability for the particle to lie in
$\Delta$ undisturbed. Indeed, one way to determine whether the particle
is in $\Delta$ is simply to measure the position at $t$ precisely.

The author has always found this distinction between
types of alternatives confusing. He did not doubt Feynman's ability
``to tell what kind of alternative is involved'', but he was less sure
of his own. This was especially the case since the distinction seemed to 
involve analyzing, not only the particular experiment 
in question, but also many others that {\it might} have been carried out. 
No 
precise rules for analyzing a given experimental situation seemed to be 
available. This situation is considerably clarified in the quantum
mechanics of closed systems.

In the quantum mechanics of closed systems, we cannot have a fundamental
distinction between ``interfering'' and ``non-interfering'' alternatives
based on different types of measurement situations, because alternatives 
are not necessarily associated with measurement situations. Whether
alternatives interfere with one another, or do not,
depends on the boundary conditions and Hamiltonian that define the 
closed system. A quantitative measure for the degree of interference is
provided by the dechoherence functional. To illustrate this idea, let us
consider the single particle model we have been discussing on the time 
interval $[t_0,t]$. The fine-grained histories are the particle paths on
this interval. Sets of alternatives correspond to partitions of these
paths into an exhaustive set of exclusive classes $\{c_\alpha\}, \>
\alpha=1, 2, \cdots$. The classes are coarse-grained alternatives for 
the closed system. For example, one
could partition the paths by which of an exhaustive set of position
intervals they pass through at one time, which of a different set of
position intervals they pass through at another time, etc. There are
many more general possibilities (see, e.g. Hartle, 1991). The
decoherence functional is a complex valued functional on pairs of 
coarse-grained alternatives defined in a sum-over-histories
formulation of the quantum mechanics of a closed system by:
$$
D(\alpha', \alpha)=N \int\nolimits_{c_{\alpha'}} \delta x'
\int\nolimits_{c_\alpha} \delta x\ \rho_f(x_f,x^\prime_f) {\rm exp}
\left\{i\left(S[x'(\tau)] - S[x(\tau)]\right)/\hbar \right\} \rho_i(x'_0,x_0). 
\tag fiveten
$$
The first sum is over paths $x'(t)$ in the class $c_{\alpha'}$ and
includes a sum over their initial endpoints $x'_0$ and final endpoints
$x'_f$. The sum over paths $x(t)$ is similar. The normalization
factor is $N=1/Tr(\rho_f\rho_i)$ where the $\rho$'s are the operators
whose matrix elements appear in \(fiveten).
We have written the decoherence functional for a general, time-neutral,
 formulation of quantum
mechanics\footnote*{ See, e.g. Aharonov, Bergmann and Lebovitz (1964)
in the quantum mechanics of measured subsystems, and Griffiths (1984)
and Gell-Mann and Hartle (1993) in the quantum mechanics of closed
systems. } in which both an initial and a final condition enter 
symmetrically,
  represented by 
density matrices $\rho_i(x'_0, x_0)$ and  $\rho_f(x'_f, x_f)$
respectively. 
 The final condition which seems to best represent our
universe and ensures causality is a final condition of indifference
with respect to final state 
in which the final density operator is $\rho_f \propto I$. 

The ``off-diagonal'' elements of the decoherence functional $(\alpha'
\not=\alpha)$
are a measure of
the degree of interference between pairs of alternatives. When the
interference is negligible between all pairs in an exhaustive set, 
the ``diagonal'' elements
$(\alpha'=\alpha)$ are the probabilities of the alternatives
 and obey the correct
probability sum rules as a consequence of the absence of interference.  
The orthogonality of the branches
in \(twonine) is an operator transcription of this condition in the special
case that $\rho_f \propto I$.

The important point for a discussion of ``interfering'' and
``non-interfering'' alternatives is that {\it all} alternatives are
potentially interfering in the quantum mechanics of closed systems. For this reason amplitudes are summed over them
in the construction of the decoherence functional
 \(fiveten). Whether alternatives are interfering or not
depends on the measure of interference provided by \(fiveten), but in
its construction all sets of alternatives are treated the same. 
When interference between each pair is negligible
 the probabilities for coarser-grained
alternatives may be constructed either directly from \(fiveten) by
summing amplitudes, or 
by summing the probabilities for the finer-grained alternatives in the
coarser-grained ones.  The equivalence between the two is the 
content of decoherence.             

Thus, there is no distinction between kinds of alternatives generally in
the formalism, but
 distinctions may emerge between different kinds of alternatives
because of particular properties of $\rho_i$
and $\rho_f$. In particular, if $\rho_f \propto I$ any
alternatives at the last time will decohere. Thus,
indifference with respect to final states is, in a time-neutral
formulation of quantum mechanics, the origin of the
usual rule that final alternatives are ``non-interfering'' rather than
an analysis of whether ``one could have measured them but didn't''.

\vskip .26 in
\centerline{\bf VI. Conclusion}
\vskip .13 in
In the quantum mechanics of closed systems, projections act on states in
the formula for the probabilities of histories, but those reductions are
not necessarily associated with a measurement situation within the
system and certainly not with one from without. Uncertainty principles
limit what kinds of alternatives a set of projections can describe, but these 
limitations need not be of our ability to carry out a
 measurement.  Interfering
alternatives can be distinguished from non-interfering ones, not by
analyzing what might have been measured, but by using the decoherence
functional as a quantitative measure of interference.  Probabilities can
be consistently assigned only to non-interfering sets of alternative
 histories but
decoherence as a consequence of a particular initial condition and
Hamiltonian rather than measurement decides which sets these are.

The fundamental role played by measurement in formulating a quantum
mechanics of subsystems is replaced by decoherence in the quantum
mechanics of a closed system.  In the opinion of the author, the result is not
only greater generality so that the theory can be applied to
cosmology, but also greater clarity.  An important reason for this is
the disassociation of the notion of alternative from an ideal
measurement.  As we saw from the work of Wigner, Araki and Yanase, and
the argument of Section IV, ideal measurements are almost impossible to
realize exactly within quantum mechanics, and are therefore of limited
value as approximations to realistic measurement situations.  But
in the usual quantum mechanics of measured subsystems, the second
law of evolution is stated for ideal measurements, not
realistic ones.  To discuss the evolution under realistic alternatives
it appears necessary to consider more and more of the universe beyond
the subsystem of interest until one obtains a subsystem large enough
such that measurements of {\it it} may be approximated as ``ideal''.  By
contrast, the alternatives used in the quantum mechanics of closed
systems are general enough to describe realistic measurement situations.
The theory can provide quantitative estimates of their closeness to
``ideal'' and therefore to how closely
the quantum mechanics of measured subsystems approximates 
the more general quantum mechanics of closed systems.

Thus, little havoc needs be made to achieve a quantum mechanics of
closed systems.  All that is needed is a more general formulation in
which decoherence rather than measurement is fundamental, but in which
most features of the approximate quantum mechanics of subsystems
that were tied to measurement re\"emerge in a more general and
 conceptually clearer light.
\vskip .26 in
\centerline{\bf Acknowledgment}
\vskip .13 in
The work described in this essay is an outgrowth of the author's program
with M.~Gell-Mann to explore and clarify quantum mechanics.  It is a
pleasure to thank him for many discussions that are reflected, in part,
in the work presented here.

\vfill\eject

\singlespace
\gdef\journal#1, #2, #3, 1#4#5#6{(1#4#5#6) {\sl #1} {\bf #2}, #3}
\beginparmode\frenchspacing
               \leftskip=1truecm\parskip=8pt plus 3pt
                \everypar{\hangindent=3em}
\vskip .26 in
\centerline{\bf References}
\vskip .13 in
\noindent Aharonov, Y., Bergmann, P.,  and Lebovitz,  J. \journal Phys.
Rev.,
B134, 1410, 1964.

\noindent Araki, H.~and Yanase, M. \journal Phys. Rev., 120, 622, 1960.

\noindent Brill, D.~and Hartle, J.B. \journal Phys. Rev. B, 135, 271,
1964.

\noindent Busch, P. \journal Found. Phys., 17, 905, 1987.

\noindent Caves, C.  \journal Phys. Rev. D, 33, 1643, 1986.

\noindent \underbar{~~~~~~~~~~}   \journal Phys. Rev. D, 35, 1815, 1987.

\noindent DeWitt, B. \journal Physics Today, 23, no. 9, 1970.

\noindent Everett, H. \journal Rev. Mod. Phys., 29, 454, 1957.

\noindent Feynman,  R.P. and Hibbs, A. (1965) {\sl Quantum Mechanics and
Path
Integrals}, McGraw-Hill, New York.

\noindent Gell-Mann, M.~and Hartle, J.B. (1990a) in {\sl Complexity,
Entropy,
and the Physics of Information, SFI Studies in the Sciences of
Complexity}, Vol.
VIII, ed. by W. Zurek,  Addison Wesley, Reading or in {\sl Proceedings
of
the 3rd
International Symposium on the Foundations of Quantum Mechanics in the
Light of
New Technology} ed.~by S.~Kobayashi, H.~Ezawa, Y.~Murayama,  and
S.~Nomura,
Physical Society of Japan, Tokyo.

\noindent  Gell-Mann, M.~and Hartle, J.B. (1990b)
in the {\sl Proceedings of
the 25th International Conference on High Energy Physics, Singapore,
August,
2-8, 1990},
ed.~by K.K.~Phua and Y.~Yamaguchi (South East Asia Theoretical
Physics Association
and Physical Society of Japan) distributed by World Scientific,
Singapore.

\noindent Gell-Mann, M.~and Hartle, J.B. (1993a) {\it Classical Equations
for Quantum Systems} 
 (to be published in {\sl Phys.~Rev.~D}).

\noindent Gell-Mann, M. and Hartle, J.B. (1993b) in {\sl Proceedings of the
1st International A. D. Sakharov Conference on Physics, USSR, May 27-31,
1991} and in {\sl Proceedings of the NATO Workshop on the Physical
Origins of Time Assymmetry, Mazagon, Spain, September 30-October4, 1991}
ed. by J. Halliwell, J. Perez-Mercader, and W. Zurek, Cambridge
University Press, Cambridge.

\noindent Griffiths, R.  \journal J. Stat. Phys., 36, 219, 1984.

\noindent Groenewold, H.J. (1952) {\sl  Proc. Akad. van Wetenschappen},
Amsterdam, Ser. B, {\bf 55}, 219.

\noindent Hartle, J.B. (1991a) {\it The Quantum Mechanics of Cosmology},
in {\sl
Quantum
Cosmology and Baby Universes:  Proceedings of the 1989 Jerusalem Winter
School for Theoretical Physics}, ed. by ~S.~Coleman, J.B.~Hartle,
T.~Piran,
and S.~Weinberg, World
Scientific, Singapore,  pp. 65-157.

\noindent Hartle, J.B. \journal Phys. Rev. D, 44, 3173, 199{1b}.

\noindent Hartle, J.B. (1993a) {\it The Quantum Mechanics of Closed systems} in
the {\sl Festschrift for C.W.~Misner}, ed.~by B.-L. Hu, M.P.~Ryan, and
C.V.~Vishveshwara, Cambridge University Press, Cambridge.

\noindent Hartle, J.B. (1993b) {\it Spacetime Quantum Mechanics and the
Quantum Mechanics of Spacetime} in {\sl Proceedings of the 1992 Les
Houches
Summer School Gravitation and Quantizations}, ed.~by B.~Julia, North
Holland, Amsterdam. 

\noindent Isaacson, R. \journal Phys. Rev., 166, 1263, 196{8a}.

\noindent \underbar{~~~~~~~~~~} \journal Phys. Rev., 166, 1272, 196{8b}.

\noindent London, F. and Bauer, E. (1939) {\sl La th\'eorie de
l'observation en
m\'ecanique quantique}, Hermann, Paris.

\noindent Margenau,  H.  \journal Phil. of Sci., 25, 23,
1958.

\noindent Omn\`es, R. \journal J. Stat. Phys., 53, 893, 198{8a}.

\noindent \underbar{~~~~~~~~~~} \journal J. Stat. Phys., 53, 933,
198{8b}.

\noindent \underbar{~~~~~~~~~~}
\journal J. Stat. Phys., 53, 957, 198{8c}.

\noindent Page, D. and Wootters, W. \journal Phys. Rev. D, 27, 2885, 1983.

\noindent Prugov\v cki, E. \journal Can. J. Phys., 45, 2173, 1967.

\noindent  Stachel,  J. (1986) in {\sl From Quarks to Quasars}, ed.~by
R.G.~Colodny, University of Pittsburg Press, Pittsburgh,  p. 331ff.

\noindent von Neumann,  J. (1955) {\sl Mathematische Grundlagen der
Quantenmechanik}, J. Springer, Berlin (1932). [English trans. {\sl
Mathematical
Foundations of Quantum Mechanics}, Princeton University Press,
Princeton].

\noindent Wheeler, J.A. (1964) in {\sl Relativity Groups and Topology},
ed.~by B.~DeWitt and C.~DeWitt, Gordon and Breach, New York.

\noindent Wigner, E.P. \journal Zeit. f. Phys., 131, 101, 1952.

\noindent \underbar{~~~~~~~~~~} \journal Am. J. Phys., 31, 6, 1963.

\endit
\end